\documentclass[10pt,letterpaper]{article}
\usepackage[top=0.85in,left=2.75in,footskip=0.75in]{geometry}

\usepackage{amsmath,amssymb}

\usepackage{changepage}

\usepackage{textcomp,marvosym}

\usepackage{cite}

\usepackage{nameref,hyperref}

\usepackage[right]{lineno}

\usepackage[nopatch=eqnum]{microtype}
\DisableLigatures[f]{encoding = *, family = * }

\usepackage[table]{xcolor}

\usepackage{array}

\newcolumntype{+}{!{\vrule width 2pt}}

\newlength\savedwidth

\raggedright
\usepackage[aboveskip=1pt,labelfont=bf,labelsep=period,justification=raggedright,singlelinecheck=off]{caption}

\makeatletter
\renewcommand{\@biblabel}[1]{\quad#1.}
\makeatother

\usepackage{lastpage,fancyhdr,graphicx}
\usepackage{epstopdf}
\fancyheadoffset[L]{2.25in}
\fancyfootoffset[L]{2.25in}
\begin{document}
\vspace*{0.2in}

\begin{flushleft}
{\Large
\textbf\newline{Memory signatures in path curvature of self-avoidant model particles are revealed by time delayed self mutual information} % Please use "sentence case" for title and headings (capitalize only the first word in a title (or heading), the first word in a subtitle (or subheading), and any proper nouns).
}
\newline
\\
Katherine Daftari\textsuperscript{1} and
Katherine A.~Newhall\textsuperscript{1*}
\\
\bigskip
\textbf{1} Department of Mathematics, University of North Carolina at Chapel Hill, Chapel Hill, North Carolina, USA
\\
\bigskip

% Insert additional author notes using the symbols described below. Insert symbol callouts after author names as necessary.
% 
% Remove or comment out the author notes below if they aren't used.
%
% Primary Equal Contribution Note
%\Yinyang These authors contributed equally to this work.

% Additional Equal Contribution Note
% Also use this double-dagger symbol for special authorship notes, such as senior authorship.
%\ddag These authors also contributed equally to this work.

%\textcurrency Current Address: Dept/Program/Center, Institution Name, City, State, Country % change symbol to "\textcurrency a" if more than one current address note
% \textcurrency b Insert second current address 
% \textcurrency c Insert third current address


% Group/Consortium Author Note
%\textpilcrow Membership list can be found in the Acknowledgments section.

% Use the asterisk to denote corresponding authorship and provide email address in note below.
* knewhall@unc.edu

\end{flushleft}
\section*{Abstract}
Emergent behavior in active systems is a complex byproduct of local, often pairwise, interactions. 
One such interaction is self-avoidance, which experimentally can arise as a response to self-generated environmental signals; such experiments have inspired non-Markovian mathematical models.
In previous work, we set out to find ``hallmarks of self-avoidant memory" in a particle model for environmentally responsive swimming droplets.
In our analysis, we found that transient self-trapping was a spatial hallmark of the particle's self-avoidant memory response.
The self-trapping results from the combined effects of behaviors at multiple scales: random reorientations, which occur on the diffusion scale, and the self-avoidant memory response, which occurs on the ballistic (and longer) timescales. 
In this work, we use the path curvature as it encodes the self-trapping response to estimate an ``effective memory lifetime" by analyzing the decay of its time-delayed mutual information and subsequently determining the longevity of significant nonlinear correlations. 
This effective memory lifetime (EML) is longer in systems where the curvature is a product of both self-avoidance and random reorientations as compared to systems without self-avoidance.

% Please keep the Author Summary between 150 and 200 words
% Use first person. PLOS ONE authors please skip this step. 
% Author Summary not valid for PLOS ONE submissions.   
%\section*{Author summary}
%Lorem ipsum dolor sit amet, consectetur adipiscing elit. Curabitur eget porta erat. Morbi consectetur est vel gravida pretium. Suspendisse ut dui eu ante cursus gravida non sed sem. Nullam sapien tellus, commodo id velit id, eleifend volutpat quam. Phasellus mauris velit, dapibus finibus elementum vel, pulvinar non tellus. Nunc pellentesque pretium diam, quis maximus dolor faucibus id. Nunc convallis sodales ante, ut ullamcorper est egestas vitae. Nam sit amet enim ultrices, ultrices elit pulvinar, volutpat risus.

%\linenumbers

% Use "Eq" instead of "Equation" for equation citations.
\section*{Introduction}
\label{sec:Intro}
Active systems are comprised of one or many individual living or nonliving units that harness energy to produce mechanical work used for locomotion. 
In nature, such systems span from the microscale to the macroscale. 
On the microscale, bacterial colonies have been observed to respond to environmental cues, such as local chemical gradients \cite{chemotaxis_living_pioneer_1,chemotaxis_living_pioneer_2,chemotaxis_living_recent_1}, gravity \cite{gravitaxis_living_pioneer_1,gravitaxis_living_recent_1}, and sources of light \cite{photoelectric_living_pioneer_1, photoelectric_living_pioneer_2,  photoelectric_living_recent_2}.
Shoaling \cite{fish1,fish2, goby}, swarming \cite{honeybees, locusts, fireflies}, flocking \cite{starlings, starlings2}, and herding \cite{sheep} are all examples of emergent behavior in macroscale living systems which have been studied extensively.
These behaviors are believed to serve evolutionary purposes such as protection from predators \cite{fish2,sheep} and more efficient foraging \cite{sheep}.
Many inventive nonliving systems take inspiration from these biological active systems;
when such systems exist at the microscale, we refer to them as active particles that are self-propelled and are often subject to random fluctuations. 
Such self-propelled particles include (but are not limited to) autophoretic swimming droplets \cite{autophoresis_synthetic_pioneer_1, Brujic2_autophoresis_synthetic_recent_1, Maass1_autophoresis_synthetic_recent_2, Brujic1_autophoresis_synthetic_recent_3, Maass2_autophoresis_synthetic_recent_4}, chemically propelled droplets \cite{chemotaxis_synthetic_pioneer_1, chemotaxis_synthetic_pioneer_2, chemotaxis_synthetic_recent_1,chemotaxis_synthetic_recent_2}, and even light sensitive particles \cite{photoelectric_active_pioneer_1,photoelectric_active_pioneer_2,photoelectric_active_recent_1,photoelectric_active_recent_2}. 
(For a comprehensive review of micro-scale active systems and current research developments, see Refs.~\cite{Zhang, Ebbens, Ebbens_Howse}.)
Non-microscopic systems of autonomous robots or hexbugs have also been studied \cite{robots1,robots2,robots3}.

Mathematical models of such biological and synthetic active systems are often agent-based.   
 A frequently used model for an individual agent is an active Brownian particle (ABP) that prescribes a constant velocity in a slowly diffusing direction (see \cite{Active_Brownian_review} for a review of active Brownian type models).
 Interaction rules between agents, including themselves, can be specifically prescribed (i.e., agent realignment \cite{Vicsek}) or they can arise from the evolution of some modeled physical process (i.e. chemical gradient sensing \cite{Lowen_2}).
In addition, agents can be subject to random forces, such as thermal fluctuations or random reorientations. 
In our work \cite{our_paper} we studied a theoretical model of microscale swimming particles that is physically inspired by experimental droplets that ``swim" in a surfactant bath due to interactions with self-created chemical gradients. 
Dissolution of these droplets creates changing local concentration gradients that then create heterogeneity in the surface tension, inducing microcurrents called Marangoni flows.
These microcurrents propel the droplets over short distances in a ballistic fashion while local flow instabilities cause spontaneous direction changes \cite{autophoresis_synthetic_pioneer_1, Brujic2_autophoresis_synthetic_recent_1, Maass1_autophoresis_synthetic_recent_2, Brujic1_autophoresis_synthetic_recent_3, Maass2_autophoresis_synthetic_recent_4}.
Because these particles swim fast relative to the speed at which diffusion erases their self-created chemical gradients, they avoid each other's and their own past locations.  
We use the term self-avoidant to refer to this chemorepulsive behavior of a single particle.

To mimic this behavior, our simulated particles navigate the changing chemical environment by descending their self-created chemical gradients towards regions of lowest chemical density.
(For a more thorough explanation of this model and its dynamics, see \cite{our_paper}.)
The changing chemical environment $c(\mathbf{y},t)$ evolves under the diffusion equation 
\begin{subequations} \label{eq:nondim coupled}
\begin{equation} \label{eq:nondim coupled PDE}
\partial_t c(\mathbf{y},t)  = \mu\Delta c(\mathbf{y},t) +\mu \phi \exp \left[ -\frac{|\mathbf{y}-\mathbf{Y}(t)|^2}{2}\right], \quad  \mathbf{y} \in \Omega, \quad t \geq 0.
\end{equation}
with diffusion coefficient $\mu$.
Rather than explicitly accounting for the particle boundary, we model the space ``occupied" by our particle using a Gaussian scaled by $\mu\phi$ for the source term.
We evolve the noisy particle's position $\mathbf{Y}(t)$ by the overdamped Langevin equation 
\begin{equation} \label{eq:nondim coupled SDE}
d\mathbf{Y}(t)  = - \nu \left(\int_{\Omega}\exp \left[ -\frac{|\mathbf{y}-\mathbf{Y}(t)|^2}{2}\right] \nabla_{\mathbf{y}} c (\mathbf{y},t)d \mathbf{y}\right)dt   +\sqrt{\epsilon} d\mathbf{B}(t).
\end{equation}
\end{subequations}
The deterministic integral term scaled by strength $\nu$ calculates the overall effect of the chemical gradient on the particle; as in Eq.~\eqref{eq:nondim coupled PDE} the point particle's spatial footprint is modeled by a scaled Gaussian. 
We note here that the negativity of this integral term mathematically produces the gradient descent that induces self-avoidance.
Positivity of this term would make the particle self-attracting, a case that was studied in \cite{Lowen_2, Grima, Golestanian}.
The noise is modeled with a Wiener process, $\mathbf{B}(t)$ scaled by $\epsilon$.

By solving the diffusion PDE \eqref{eq:nondim coupled PDE} and explicitly calculating the gradient term in Eq.~\eqref{eq:nondim coupled SDE}, our model is reduced to the SDE 
\begin{equation} \label{eq:model}
 d\mathbf{Y}  =   \frac{\pi}{2}\mu \nu \phi \int_{0}^{t}\exp \left[ -\frac{|\mathbf{Y}(t)-\mathbf{Y}(s)|^2}{4(1+ \mu(t-s))}\right] 
    \frac{\mathbf{Y}(t)-\mathbf{Y}(s)}{(1+ \mu(t-s))^{2}}ds  dt + \sqrt{\epsilon} d\mathbf{B}.
\end{equation}
for the path dynamics.  
In this form, the non-Markovian aspect of the model is more apparent, since the deterministic component integrates the location of the particle over all past times up to the current time $t$.
Additionally, this form highlights the role of the diffusion coefficient $\mu$ in influencing the memory of the model as $\mu$ appears in the memory kernel.  
An important observation from our previous work \cite{our_paper} was the inability to \emph{independently} tune the memory, $\mu$, and the dynamic regime of the model.
Larger $\mu$, corresponding to faster diffusion, leads to a more quickly decaying kernel, thereby diminishing the effect of past history on the future evolution of the particle.
The self-created gradient then becomes too small and ballistic swimming is lost, resulting in (non-active) Brownian motion.
By decreasing $\mu$ to add more memory, the droplet's source strength is also diminished, similarly killing the ballistic swimming and resulting in Brownian motion.
Therefore there is an intermediate regime of $\mu$ values that permits both ballistic swimming and random reorientations that allow for self-interactions and thereby the self-trapping effect of memory.

\begin{figure}[!h]%
\centering
\includegraphics[width=\textwidth]{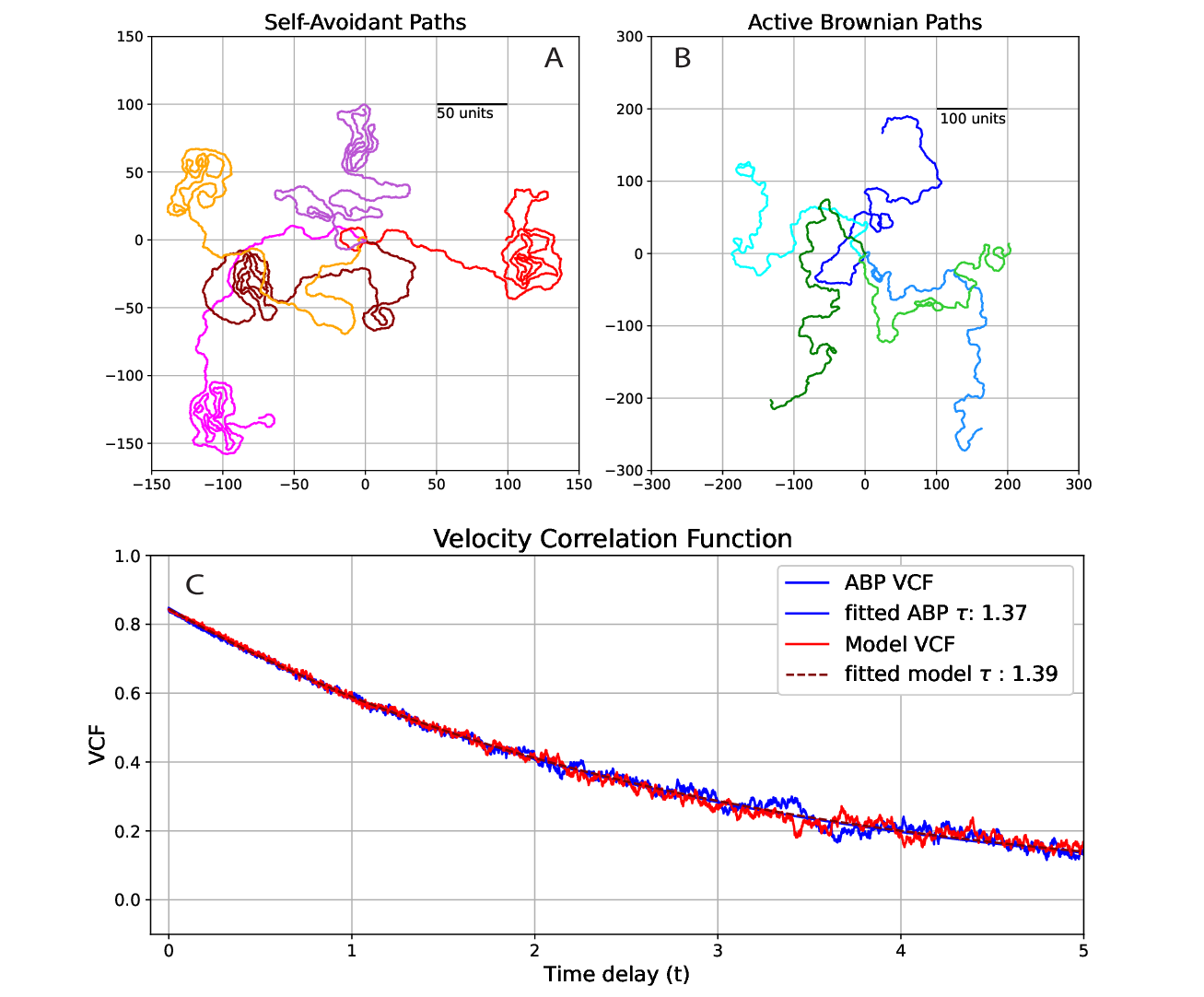}%
\caption{\emph{(A) Five independent self-avoidant paths for comparison to (B) five independent active Brownian particle paths with matching velocity $V$ and reorientation timescale $\tau$ derived from fitting the velocity correlation function shown in (C).
Despite nearly identical VCFs, the paths are qualitatively different.
}
}
\label{fig:Path compare and VCF compare}
\end{figure}

The previous study of our model in \cite{our_paper} showed what we think to be the first instance of \textit{transient self-trapping} in a self-avoidant model.  
Self-trapping begins with random reorientations that initiate interactions with the particle's past history.
The particle's self-avoidant memory then causes its trajectory to temporarily spiral in on itself in an effort to avoid visiting the past path locations [see Fig. \ref{fig:Path compare and VCF compare}(A)]. 
For a self-attracting particle, self-trapping has been seen experimentally before \cite{Liebchen_and_Lowen, Tsori}; only recently has trapping of chemo-repulsive particles been shown experimentally \cite{Maass_new}.

This counter-intuitive behavior for a self-avoidant particle produces path segments of extremely high curvature and comparatively dense path data. 
These signatures of trapping in the path data do not show up in existing metrics that completely characterize ABP-like motion, namely the mean square displacement (MSD) and the orientation decorrelation timescale $\tau$.  
For the first example, the MSD curve of the self-avoidant particles exhibits the 3 classic displacement regimes (diffusion at small timescales, transition to ballistic at intermediate timescales, and then to enhanced diffusion at long time scales) and is well fit by the theoretical curve for active Brownian particles (ABP).
The temporal averaging of the MSD does not permit distinction between the self-avoidant particles that  transiently switch between two regimes, self-trapping and free motion, and an ABP system with appropriate parameters.

Another insufficient existing statistic correlated to curvature for active particles is the orientation decorrelation timescale $\tau$, which measures the rate at which a particle becomes decorrelated with, or ``forgets", its past movement direction.
In the ABP model, the value of $\tau$ can be determined directly from the model equations as the rotational noise strength is explicitly specified; i.e.~if the rotational diffusion of ABP is given by $d\theta = \tau^{-1/2} dB_{\theta}$, then the instantaneous velocity vector correlation function (VCF) takes the form $e^{-\frac{t}{2 \tau}}$. 
In contrast, for systems like ours where $\tau$ cannot be determined analytically, it can only be estimated empirically by fitting the VCF with an exponential.

As with the MSD, we have found $\tau$ to be limited in its explanatory power over the curvature of paths in our self-avoidant model.
The decorrelation timescale $\tau$ is dominated by random reorientations in the ballistic direction over the longer-time deterministic effects of the self-avoidant memory on the path curvature.
To reinforce this point, in Fig.~\ref{fig:Path compare and VCF compare}, we compare ABP paths generated with $\tau$ fitted from the VCF of the self-avoidant model, and velocity $V$ chosen to match the self-avoidant paths. 
(The velocity $V$ can be calculated analytically as a function of the parameters $\mu$, $\nu$, and $\phi$ for the self-avoidant model \cite{our_paper}.)
Despite identical generating parameters ($V$ and $\tau$) in addition to closely matching VCF curves, we observe noticeable differences in the trajectories, particularly the curvature.
In fact, self-trapping (and the resulting curvature) is a mesoscale \textit{emergent} effect of self-avoidant memory that arises organically as the system evolves and creates qualitative changes in the curvature over longer timescales than a random reorientation.
Therefore, it is unsurprising that a single reorientation timescale that captures the decay of \textit{linear} correlations between consecutive orientations does not capture the effects of a self-avoidant memory response. 

For alternative ways of detecting the extent to which knowledge of the past history influences the future behavior of the system, one might consider information theory, which studies the communication of information in the presence of uncertainty.
In particular, it provides tools to identify and quantify causal relationships between signals.
Transfer entropy is one such tool that is favored for isolating directional interactions in coupled systems \cite{bats1, bats2, swallows2, zebrafish, Ben}; other information theoretic approaches include conditional mutual information \cite{Conditional_MI}, partial mutual information \cite{Partial_MI}, causation entropy \cite{Sun_causation_entropy} (and the related optimal causation entropy \cite{Sun_OCSE}), momentary information transfer \cite{momentary_info_transfer}, and Granger causality \cite{Granger}.
(Reviews include \cite{TS_review2}).
For the purpose of detecting directional interactions in coupled systems, transfer entropy-based methods \textit{eliminate} the influence of the target variable's past history (by conditioning on it) to isolate the additional influence of the source variable. 
In contrast, we aim to {\em illuminate} the non-Markovian cumulative influence of the past history of a \textit{single} particle on itself. 
Therefore, we adapt the mutual information, which quantifies the ``distance from independence" between two random variables (in our case, a variable with its own past history) to capture this self-interaction between the present and past.
Since the high-curvature regions of self-trapping are an emergent spatial hallmark of self-avoidant memory, we use the temporal structure of the time-lagged path curvature mutual information to reveal the presence of past-history effects on the self-avoidant particles when compared to Markovian ABPs. 

Our paper is structured as follows. 
We begin in Methods by summarizing a method for calculating the relative straightness of path data.
Following, we review the basics of mutual information and our adapted sampling method for minimizing inherent correlations in continuous time series data. 
In Effective Memory Lifetime, we present self time-delayed mutual information at various delay times, arriving at a single statistic associated with the presence of self-trapping and the overall memory level in the system.
We end the paper with conclusions.

\section*{Methods}
\label{sec:Methods}

% For figure citations, please use "Fig" instead of "Figure".
% Place figure captions after the first paragraph in which they are cited.
%\begin{figure}[!h]
%\caption{{\bf Bold the figure title.}
%Figure caption text here, please use this space for the figure panel descriptions instead of using subfigure commands. A: Lorem ipsum dolor sit amet. B: Consectetur adipiscing elit.}
%\label{fig1}
%\end{figure}

The curvature due to self-trapping in our model is a response generated by the self-avoidant memory.  
To detect this memory, we quantify the correlations in the time-evolving curvature as the trajectory switches between high-curvature self-trapping and straighter, active Brownian-like states.
We begin by computing the path curvature time series at the appropriate spatial-temporal scale and then compute the time-delayed self-mutual information of this series over various lag times. 
This requires adapting the estimation of mutual information to time series data to filter out dynamic correlations (or autodependencies) inherent in time series data.

\subsection*{Straightness Index}

\begin{figure}[!h]%
\centering
\includegraphics[width=\textwidth]{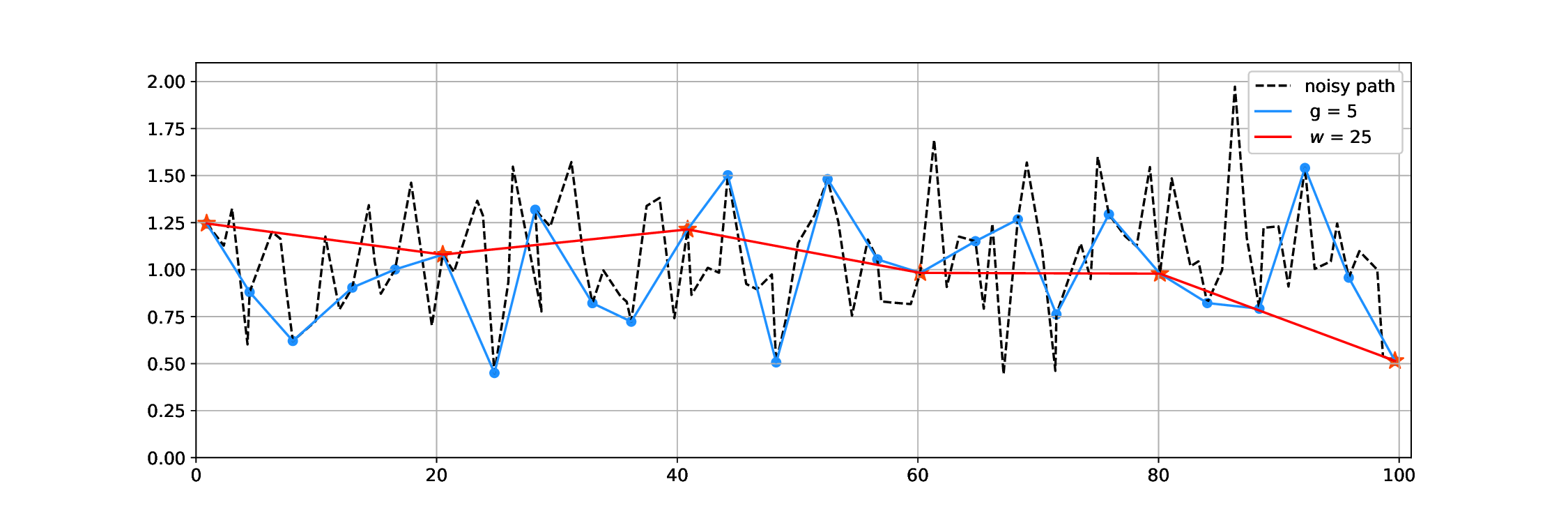}
\caption{\emph{Illustration of the straightness index computation for 
 sample noisy data (dashed gray path).  The arclength is estimated using distances at the granularity size ($g = 5$ in this example -- blue segments) while the beeline distance is estimated at the moving window size ($w=25$ in this example -- red segments).
The SI for each moving window is the ratio of its beeline to its arclength distances.
}}
\label{fig:SI schematic}
\end{figure}

Inspired by \cite{Postlethwaite}, we compute a straightness index (SI) that estimates the curvature of path data as the ratio of beeline distance to arc length.  This index allows for two relevant timescales, $g$ and $w$, to be specified  to estimate the straightness.
First, the path data is smoothed by downsampling with frequency $g$ to eliminate the effects of random noise and highlight the more deterministic path features.
Following, a moving window of size $w$ is applied to the smoothed path data.
Within this window, the beeline distance from the window start time to the window end time is computed and divided by the arc length of the path segment.
SI values close to 1 indicate a similar ratio of beeline distance to arclength, indicating straight motion and therefore low curvature.
Conversely, a low ratio of beeline distance to arclength will produce straightness values near zero and indicative of high curvature.
A schematic of the SI calculation using dummy path data is shown in Fig.~\ref{fig:SI schematic}).

Inspection of the trajectories in Fig.~\ref{fig:Path compare and VCF compare} shows that the self-trapping occurs on an intermediate spatial scale, therefore, we tune the SI to the mesoscale at which self-trapping occurs to capture the temporally changing path curvature. 
We choose $g$ to smooth out small-timescale diffusion and $w$ on the order of $V$, which is the ballistic timescale. 
We construct a time-series of SI values which functions as a time-evolving order parameter for the curvature generated by path data.
By considering individual trajectories we find that the SI captures the observed curvature well, as it varies throughout the experiment. 
In Fig.~\ref{fig:Path SI and MI} we illustrate the resultant SI time series data of three selected paths that are generated with the same model parameters, but express obviously different path features. 
The SI data (middle column) is color-coordinated to match the relevant path segment (left column).
Path C becomes self-trapped near the end of the experimental timeframe and this is reflected in the SI as repeated excursions toward values close to zero. 
Contrast this with path A, whose only excursion to zero at approximately $t = 20s$ is short-lived and is reflected in the path data as a small bend with no spiral-like pattern.
Recall it is structure or correlations in SI that we care about to detect memory, not just the existence of variable curvature.

\subsection*{Time Delayed Self Mutual Information}

In the previous section, we presented a direct way of quantifying curvature with the SI, but further work is required to directly connect the curvature to the self-avoidant memory.
Rather than the decay of linear correlations of the orientation that the VCF provides, we analyze the temporal structure of \textit{nonlinear} correlations by estimating the time-delayed self mutual information of the straightness time series data.
Mutual information was first introduced in \cite{Shannon} to assess the strength of nonlinear correlations between two random variables.
Given $X \sim p_X(x)$ and $Y \sim p_Y(y)$ with supports $\mathcal{X}$ and $\mathcal{Y}$, respectively:

\begin{equation}
   MI(X;Y)=  \int_{\mathcal{X}}\int_{\mathcal{Y}} p_{X,Y}(x,y)\log{\frac{p_{X,Y}(x,y)}{p_X(x)p_Y(y)}}dxdy.
\end{equation}

\noindent
$MI(X;Y)$ is formally the Kullback-Liebler divergence between the product of the marginal distributions, $p_X(x)\cdot p_Y(y)$ and their joint distribution, $p_{X,Y}(x,y)$. 
Intuitively, we can interpret the mutual information as characterizing a ``distance from independence" since the integral is formally zero in the case where $X$ and $Y$ satisfy $p_{X,Y}(x,y) = p_X(x)p_Y(y)$ (completely independent) and is infinite in the case where $p_X(x) = p_Y(y)$ (completely dependent); it however is not a distance in the mathematical sense.
Mutual information of $O(1)$ between samples is typically interpreted as a strong signal. 

We note that, although transfer entropy is a more common metric used for the analysis of correlations between time series data \cite{Schrieber}, mutual information is the preferred metric for our study since it has fewer limitations, both computationally and with regard to assumptions. 
Transfer entropy captures directional information flow, which is the information gain in one signal when knowledge of a second signal is added. 
Here, since we have only one signal, mutual information between a single variable in its current state with its own past states is a more natural choice.
Secondly, transfer entropy (also known as conditional mutual information) is computationally undesirable since estimation requires conditioning on all variable past states up to a chosen time delay. 
(Such conditioning requires estimation of a high-dimensional joint distribution.)
Finally, since our particles have non-Markovian memory extending back to $t=0$, we would be required to condition all the way back to $t=0$ or choose an arbitrary cutoff.

To estimate the mutual information between random variables $X$ and $Y$, we use the k-nearest neighbors algorithm described by \cite{Kraskov}, in which the authors develop an unbiased statistical estimator that takes in bivariate data of the form $Z = \{(X, Y)\}$ and assumes that  each sample, $(X_n, Y_n)$, for $ n = 1, 2,\ldots N$, is an independent realization from a stationary distribution.
Since we aim to show that the curvature of self-avoidant paths is a \textit{response} to the interactions produced by self-avoidant memory and is therefore \textit{time-dependent}, we compute the \textit{time-delayed self mutual information} of the straightness index, $MI(S(t); S(t+T))$.
Accordingly, the independent samples of the bivariate data $Z$ becomes samples at current times $\{ t_i\}$ and at future times $\{t_i +T\}$, denoted as $\{ (S(t_1), S(t_1 + T)), (S(t_2), S(t_2 + T)), \ldots , (S(t_N), S(t_N+T)) \}$.  
We see no evidence refuting the requirement of stationarity; the ensemble mean and variance do not change significantly in time (data not shown). 
We discuss needed adaptations to maximize independence between these time samples next. 

\begin{figure}[!h]
\centering
    \includegraphics[width=0.75\textwidth]{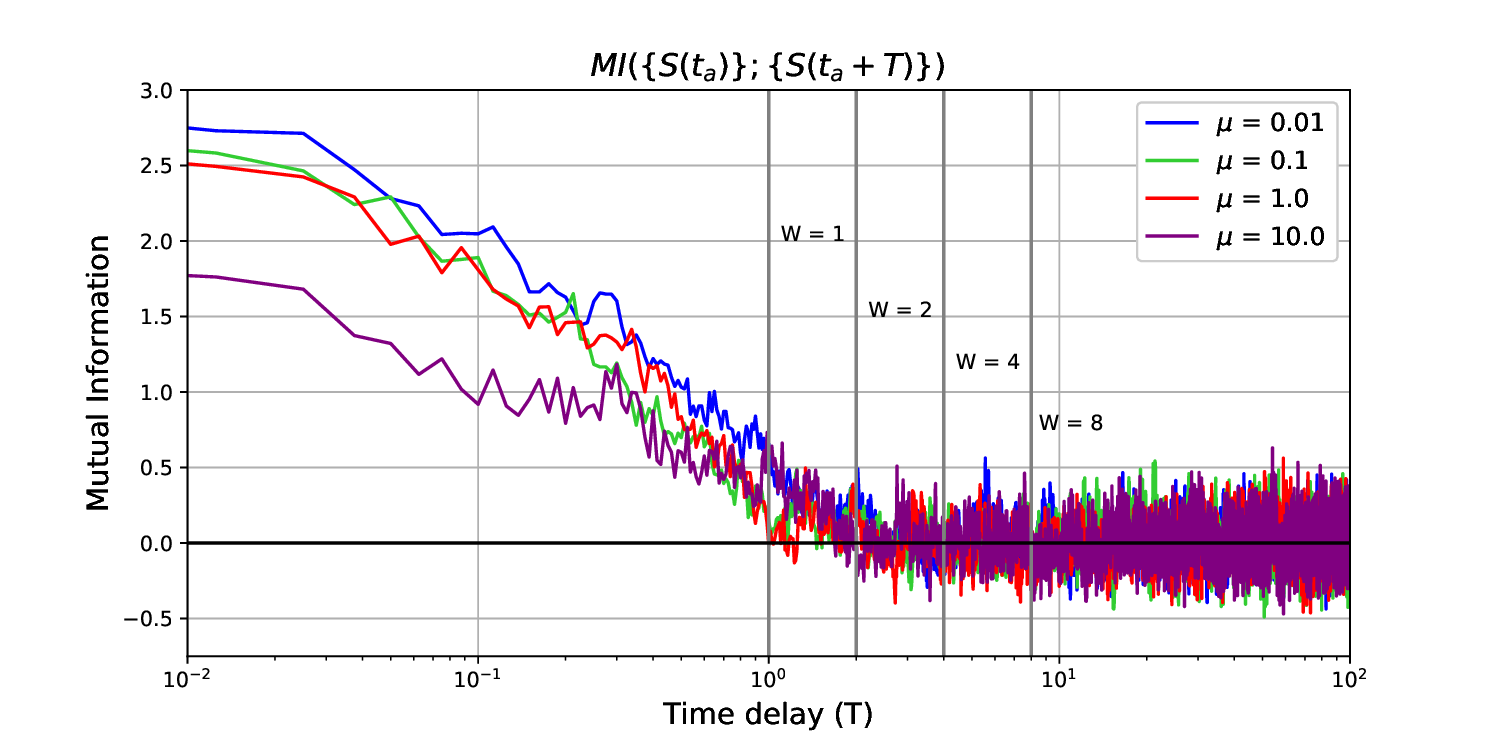}
    \caption{\emph{Estimation of nonlinear dynamical correlations in the mutual information between independently generated straightness indices across an ensemble to determine a sufficient separation window $W$ for use on a single time series to suppress these dynamical correlations.
    After some acclimation time $t_a$, we calculate the ensemble-sampled mutual information $MI(S(t_a); S(t_a + T))$, where $S(t_a)$ and $S(t_a + T)$ are the set of all straightness index values of each independently generated path at time $t_a$ and $t_a+T$. (We choose $t_a = 15$s.)
    We select $W = T$ satisfying $MI(S(t_a); S(t_a + T)) \approx 0$ as the appropriate separation window size for future use in time-sampled mutual information computations. }\label{fig:Window size calc}}
\end{figure}

One remaining issue with applying mutual information to time series data is the
introduction of correlations simply because the data is a representation of a continuous stochastic process; there are inherent autodependencies between samples that are close in time.
In contrast with how MI has been applied to time series in the literature \cite{MI_timeseries_1, MI_timeseries_2}, we utilize a technique that one of the authors developed in \cite{fish_paper}.  It is designed to suppress the effects of these dynamical correlations in the calculation of mutual information of time series data, allowing the detection of the desired self-interaction-induced correlations.
This technique enforces an average separation window, $W$, (distinct from the moving window $w$ used to calculate the SI) between consecutive sample times $t_i$ and $t_{i+1}$.  
We look for the smallest $W$ for which the mutual information of the independently generated ensemble, for which only dynamical correlations are present, is approximately zero.  
The non-zero MI seen in Figure \ref{fig:Window size calc} can only be attributed to dynamical correlations and not true history-dependent correlations.

The size of this window, which we call $W$ can be chosen as the timescale beyond which the time-delayed ensemble mutual information at any given point dips below a significance threshold (perhaps zero).
This window $W$ acts as the timescale of an effective noise filter, where the noise here is the nonzero mutual information of dynamical correlations within time-series data, which we want to exclude.
An illustration of this window size calculation for an ensemble of 96 paths is shown in Fig.~\ref{fig:Window size calc}.
We see that a window size of $W = 4$ is sufficient to filter out mutual information of dynamical correlations for the illustrative ensemble of paths.

\begin{figure}[!h]
    \includegraphics[width=1.0\textwidth]{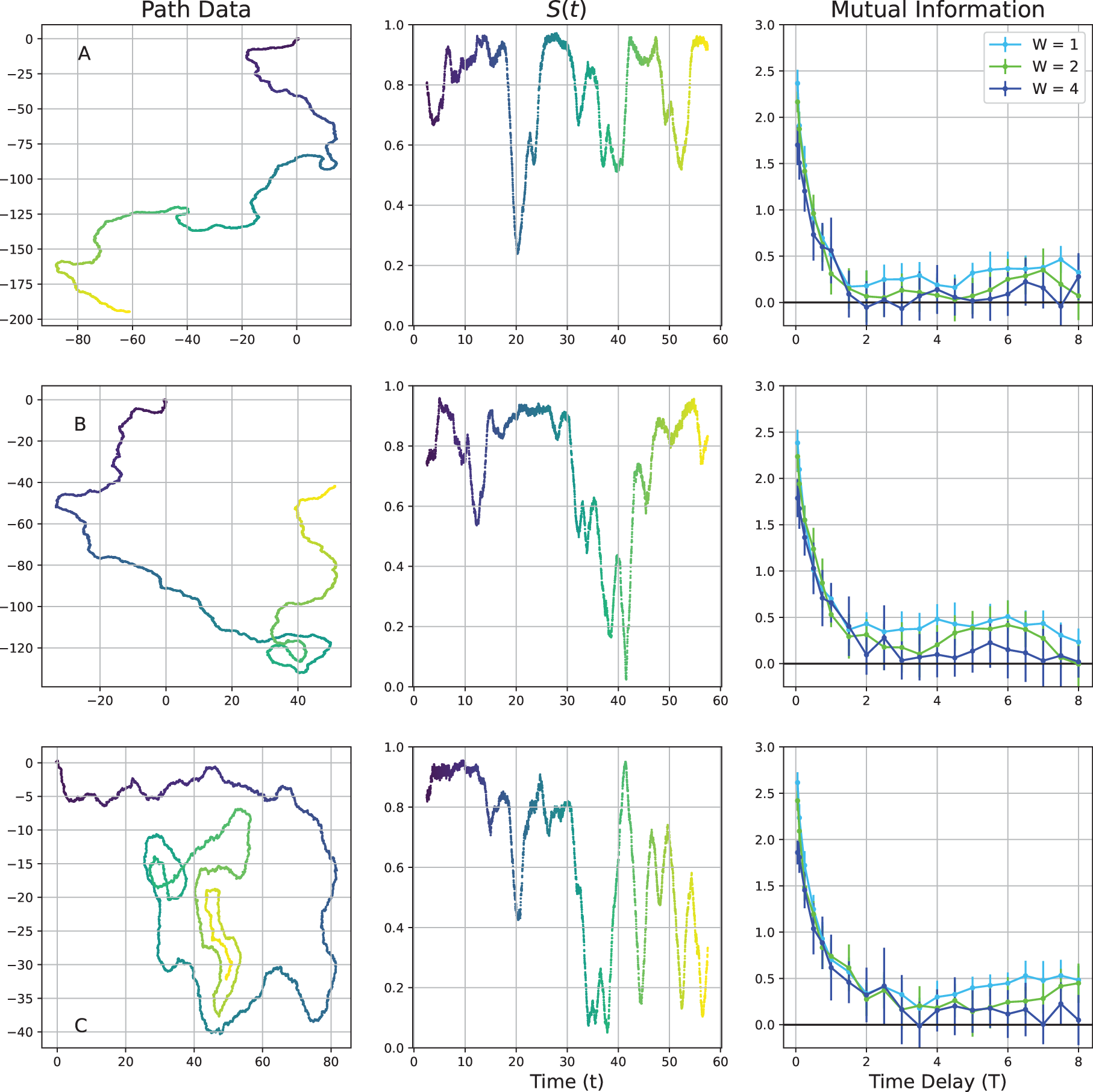}
    \caption{\emph{First column: Three 60s long paths from the same generating parameters ($\mu=0.01$ and $V=6$) that yield varying levels of self-avoidant memory expression. 
    Second column: Corresponding straightness index time series $S(t)$ for each of the three paths.
    Third column:
    The mutual information decay curves corresponding to window sizes $W = 1, 2, 4$.
    As expected, the curves only appear to decay convincingly to zero in the case that $W=4$.}
}
    \label{fig:Path SI and MI}
\end{figure}

We proceed to compute the mutual information of the straightness index for a single path as a function of the time delay $T$, $MI(S_j(t), S_j(t+T))$
We implement the method from \cite{fish_paper} for suppressing dynamical correlations
by randomly sampling $S(t)$ at times $\{t_i\} = t_0, t_1, \ldots, t_f$, satisfying $\langle t_{i+1} - t_{i} \rangle _i \approx W$.  
For the same set of model parameters, we generate three paths, $A$, $B$, and $C$, depicted in the first column of Fig.~\ref{fig:Path SI and MI}.  
For each path, we calculate the time-series of the straightness index $S_Q(t)$, $Q\in\{A,B,C\}$ shown in the second column.
Each resultant time-delayed self mutual information curve  is illustrated in the third column of Fig.~\ref{fig:Path SI and MI}, where $MI(S_A(t);S_A(t + T)$, $MI(S_B(t);S_B(t + T)$, and $MI(S_C(t);S_C(t + T)$ are shown as functions of the time delay $T$ for several values of the separation window $W$.
Confirming our choice of $W = 4$ as the dynamical correlation filtering timescale, we see that the curves associated with $W<4$ appear not to decay to zero, indicating that the dynamical correlations are still present. 
In the following section, we discuss a framework for interpreting the structure of these mutual information decay curves through the lens of the self-avoidant memory response.

% Results and Discussion can be combined.
\section*{Effective Memory Lifetime}
\label{sec:Tool}
The first observation that we make about Fig.~\ref{fig:Path SI and MI} is that the level of self-trapping in the paths, which is the self-avoidant memory expression, influences the delay time $T$ at which the mutual information becomes ``insignificant".
Path $A$, which does not become self-trapped, has mutual information that becomes insignificant the soonest, at approximately $T \approx 2$, after which it fluctuates near zero.
In contrast, nonzero mutual information of path $C$, which is self-trapped for an extended period of time (as shown in a prolonged excursion of $S_C(t)$ toward zero), persists well beyond $T \approx 2$.
Future states of the model depend on the entire spatiotemporal past history, therefore, we expect that the theoretical mutual information will never reach zero. 
However, the past influence decays in time thereby localizing the self-trapping in both space and time; we hypothesize that a finite time delay to insignificance is a good representation of the \textit{effective memory lifetime} (EML).
To definitively compare this timescale (EML) across parameters and models, we compute the first crossing of the mutual information decay curve with a chosen small, but nonzero threshold that is above the inherent fluctuations of the statistical estimator. 

\begin{figure}[!h]%
\centering
\includegraphics[width=\textwidth]{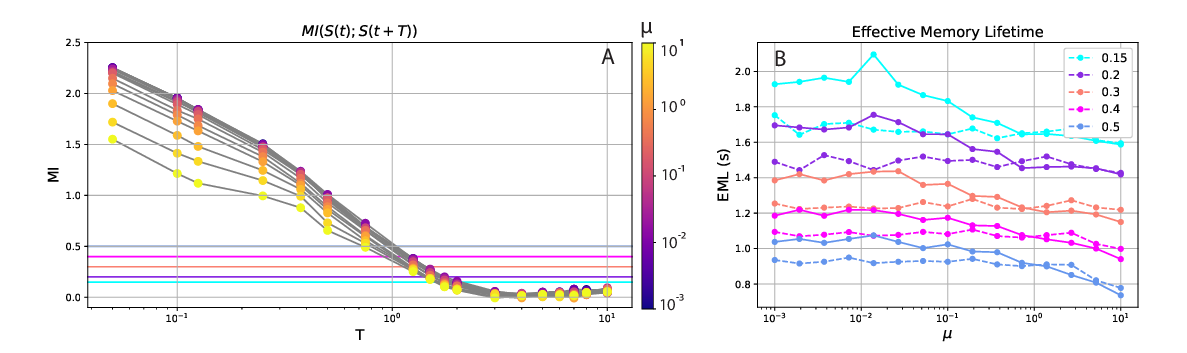}%
\caption{\emph{
(A) The average mutual information as a function of delay time decreases as the value of $\mu$ increases and the self-avoidant memory response is suppressed.
As these curves decrease in value, they cross significance thresholds (horizontal straight lines) at earlier times. 
(B) For each significance threshold, the effective memory lifetime (first crossing time) of self-avoidant paths (solid lines) is recorded as a function of $\mu$ and compared to ABP effective memory lifetimes (dashed lines) with matching $V$ and $\tau$. 
}}
\label{fig:Threshold crossing}
\end{figure}

In Fig.~\ref{fig:Threshold crossing}, we illustrate how the EML can distinguish the effect of the parameter $\mu$, which influences the expression of the self-avoidant memory through its appearance in the exponential kernel of Eq.~\ref{eq:model}.
(The velocity $V$ in the model is held constant by using the appropriate $\nu$ value for each $\mu$.)
Each curve in Fig.~\ref{fig:Threshold crossing} (A) is the average of the mutual information decay curves of 96 individual paths generated with a particular value of $\mu$ indicated by the colorbar.
As $\mu$ increases, the memory strength decreases and the corresponding mutual information decay curves become progressively lower, thus becoming insignificant sooner and having smaller EML values.

For the five threshold values chosen in Fig.~\ref{fig:Threshold crossing} (A), we compute the EML as a function of $\mu$ which are plotted in Fig.~\ref{fig:Threshold crossing} (B) (solid lines).
As $\mu$ increases (and the memory response wanes), the EML decreases irrespective of the chosen threshold value. 
For active Brownian particles (no self-avoidant memory) with matching $\tau$ and $V$, we repeat this process (dotted lines in Fig.~\ref{fig:Threshold crossing} (B)).
Irrespective of threshold value, the ABP EML are lower than those of the self-avoidant model when there is substantial self-avoidant memory (small $\mu$ values).
As the self-avoidant memory becomes weaker at larger values of $\mu$ the EML curves converge, suggesting that the self-avoidant memory response is indistinguishable from an active Brownian particle.

We explore alternative ways to control the memory by introducing a numerical solution designed to tune the effective memory; we implement this by restricting the bounds of the integral term in Eq.~\eqref{eq:model}.
The integral term represents the particle mathematically ``looking over all past times" to determine its next location.
By increasing the lower bound from zero to $\max(0,t - M)$ with $M \geq 0$, we can artificially restrict the amount of past history to affect the particle motion, and therefore reduce the effective memory.  
This restricted-memory particle model is given by
\begin{equation}
d \mathbf{Y}=\frac{\pi}{2}\mu \nu \phi \left[\int_{\max(0,t-M)}^{t}  \left ( e^{-\frac{|\mathbf{Y}(t)-\mathbf{Y}(s)|^2}{4(1 +\mu(t-s))}}\frac{\mathbf{Y}(t)-\mathbf{Y}(s)}{(1 +\mu(t-s))^{2}}\right)ds\right]dt   + \sqrt{\epsilon} d\mathbf{B}.
\label{eq:combined mod with M}
\end{equation}
(Since the combined model in Eq.~\ref{eq:model} is a non-Markovian process in which the state (position) at time $t + \delta t$ depends on all past states via the integral term,  implementation of effective memory timescale $M$ is merely limiting the number of past states that the particle has access to from $[0, t]$ to $[\max(0, t-M), t]$.)

\begin{figure}[!h]%
\centering
\includegraphics[width=3.5in]{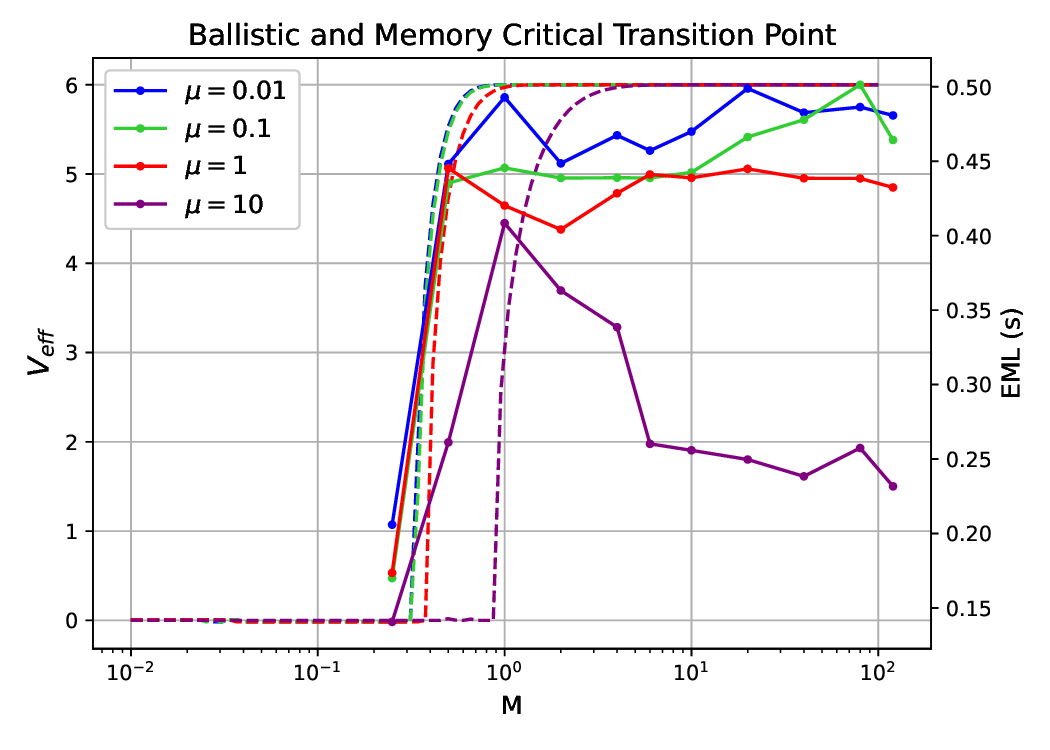}%
\caption{\emph{The velocity $V$ (dashed lines, left axis) of the truncated model Eq.~\eqref{eq:combined mod with M} is plotted together with the EML values (solid lines, right axis) as a function of the model effective memory window, $M$ (larger $M$ includes more past history in the integral).
Transition to ballistic behavior and to stable EML values occur at similar effective memory times $M$ for high and intermediate memory-inducing values of $\mu$.  
}}
\label{fig:M fig}
\end{figure}

In Fig.~\ref{fig:M fig}, we explore the effects of implementing the adapted model on the swimming velocity $V$ and the EML.
As we discuss in the introduction, the parameter $\mu$ affects both the memory and velocity of the particles, with intermediate $\mu$ values allowing for both  ballistic motion and random reorientations that are required conditions for self-trapping to arise.  
We see a similar interdependence of the parameter $M$ on both memory and velocity.
In Fig.~\ref{fig:M fig} we depict the analytical transition to non-zero velocity as a function of the effective memory window $M$ (we numerically search for a velocity $V$ so that a solution of Eq.~\eqref{eq:combined mod with M} with $\epsilon=0$ is $Y(t) = Vt$).
With the right y-axis of this same plot, we plot the EML derived from the mutual information decay curve. 
We see that the apparent transition from EML values near zero to a noisy but seemingly stable EML occurs at approximately the same critical $M$ as the nonzero velocity transition. 
The exception is that the $\mu=10$ EML does not stabilize.  In this case, with more rapid diffusion, 
the transitional $M$ is larger since more past history is needed to generate a gradient strong enough to propel the particle ballistically. 
However, once the ballistic transition has occurred, the particle velocity $V = 6$ outpaces the gradient buildup necessary to induce self-interaction.  This is shown by the EML trend back down toward zero.
We interpret this to mean that the parameter combination $V=6$ and $\mu=10$ will not yield self-trapping.
These observations further confirm that the swimming and the self-trapping memory response are intrinsically tied.

The fact that this transition between diffusion-dominated and ballistic-dominated motion is sharp removes the option of slowly tuning the effective memory response by changing $M$.
Below the critical transition value of $M$, the diffusion-dominated dynamics do not display the self-trapping memory response.  
By the time the effective velocity $V$ nearly matches the original-model velocity ($M$ order 1), the memory response has also reached its apparent full value as the path dynamics become indistinguishable as $M$ continues to increase.

\section*{Conclusion}
\label{sec:Conclusions}

We have shown how to detect memory signatures of self-avoidant model particles using the path curvature derived from trajectory data.
The path curvature is measured by a multi-scale straightness index that can be tuned to reliably capture the spatial scale of self-trapping which is an emergent response of the self-avoidant memory.
Using mutual information, we quantify nonlinear correlations in the temporal structure of the path curvature, arriving at an effective memory lifetime to capture the persistence of significant self-mutual information.

One challenge that arises when computing the mutual information from time series data, especially data derived from a continuous stochastic process, is the presence of ambient dynamical correlations.
To suppress the influence of these ambient dynamical correlations on the reported mutual information, we implement an adapted sampling scheme that enforces an average separation window between consecutive samples.
This window is chosen to be larger than the decay timescale of the dynamical correlations.
An advantage of working with model-generated data over experimental data is that we can create arbitrarily long trajectories to implement this constraint while maintaining a reasonable sample size.
Additionally, long trajectories and many replicates also increase the likelihood of observing the transient self-trapping, since it is an emergent effect of the self-avoidant memory and is not guaranteed to occur.
It may be possible to experimentally circumvent this need for long trajectories by sampling from pairwise interactions of multiple particles evolving simultaneously. 
(Note that dynamical correlations still need to be suppressed in this case.)  
Future work includes exploring this possibility for model particles in a periodic box that restricts the domain and therefore promotes more particle interactions on shorter timescales that may be more similar to the lifetime of experimental particles.

As the diffusion coefficient $\mu$ increases (thus decreasing the memory response with fixed velocity), we show that the average mutual information is lowered across all delay times and results in shorter effective memory lifetimes.  
Furthermore, we show that the EML of particles with curvature derived from self-avoidance \emph{and} random reorientations (our model) outlast the EML of particles with curvature derived \emph{only} from random reorientations (ABP with identical velocity and velocity decorrelation timescale $\tau$).
Together, these results demonstrate that the EML derived from time-delayed self-mutual information is capturing the presence of self-avoidant memory effects in trajectory data.

Further exploration of the model parameter space confirmed our first-principles arguments that the ballistic component and the self-avoidant memory response are intrinsically tied together by the diffusion coefficient $\mu$.
Because of this coupling, the memory response is not independently tunable, even if we artificially restrict the length of past history the particle accesses to be at most $M$ non-dimensional time units. 
In fact, we found a transitional $M$ beyond which model behavior changed from non-ballistic to ballistic that coincided with the stabilization of the EML.  
This further supports the entanglement of velocity and memory: the particle must move ballistically to see the self-trapping memory response and the self-avoidant memory propels the particle ballistically.

\section*{Acknowledgments}
The authors thank Daphne Klotsa and Pedro S\'aenz for feedback on an earlier version of this work.  This work is partially supported by NSF grant number DMS-2307297.

\nolinenumbers

% Either type in your references using
% \begin{thebibliography}{}
% Text
% \end{thebibliography}
%
% or
%
% Compile your BiBTeX database using our plos2015.bst
% style file and paste the contents of your .bbl file
% here. See http://journals.plos.org/plosone/s/latex for 
%\bibliographystyle{plos2015}
%\bibliography{refs}

\end{document}